\documentclass[iop, preprint]{emulateapj}
 \slugcomment{Accepted to ApJ}
\usepackage{apjfonts}
 \newcommand{\Jnic}{$J_{110}$}

 \newcommand{\iacs}{$i_{775}$}
 \newcommand{\zacs}{$z_{850}$}

 \newcommand{\lya}{Ly$\alpha$}

\begin{document}
\title{New Results from the Magellan IMACS Spectroscopic \lya\ Survey: NICMOS Observations of \lya\ emitters at $\lowercase{z}=5.7$ \altaffilmark{1} }

\author{Alaina L. Henry\altaffilmark{2}, Crystal L. Martin\altaffilmark{2}, Alan Dressler\altaffilmark{3}, Patrick McCarthy\altaffilmark{3} \&  Marcin Sawicki\altaffilmark{4}}

\altaffiltext{1}{ This work is  based in part on observations made with the NASA/ESA {\it Hubble Space Telescope}, obtained from the Space Telescope Science Institute, which is operated by the Association of Universities for Research in Astronomy Inc., under NASA contract NAS 5-26555.  These observations are associated with proposal 11183.}
\altaffiltext{2}{Department of Physics, University of California, Santa Barbara, CA 93106; ahenry@physics.ucsb.edu}
\altaffiltext{3}{Observatories of the Carnegie Institute of Washington, Santa Barbara Street, Pasadena, CA 91101}
\altaffiltext{4}{Department of Astronomy and Physics, Saint Mary's University, Halifax, NS B3H 3C3, Canada}

\begin{abstract}
 We present NICMOS \Jnic\ (rest-frame 1200-2100 \AA) observations of the three $z=5.7$ \lya\ emitters discovered in the blind multislit spectroscopic survey by  \cite{Martin08}.  These images confirm the presence of the two sources which were previously only seen in spectroscopic observations.  The third source, which is undetected in our \Jnic\ observations has been detected in narrowband imaging of the Cosmic Origins Survey (COSMOS), so our nondetection implies a rest frame equivalent width $>146$ \AA\ ($3\sigma$).    
The two \Jnic-- detected sources have more modest rest frame equivalent widths of 30-40 \AA, but all three are typical of high-redshift LAEs.   In addition, the \Jnic- detected sources have UV luminosities that are within a factor of two of $L^*_{UV}$, and sizes that appear compact ($r_{hl} \sim $0\farcs15) in our NIC2 images -- consistent with a redshift of 5.7.      
 We use these UV-continuum and \lya\ measurements to estimate the \iacs\ - \zacs\ colors of these galaxies, and show that at least one, and possibly all three would be missed by the $i-$ dropout LBG selection.     These observations help demonstrate the utility of  multislit narrowband spectroscopy as a technique for finding faint emission line galaxies.  
  
 \end{abstract}
 

\section{Introduction}
Studying the epoch around $z\sim6$ is important for our understanding of the early stages of galaxy formation.       
It is around this time, when the age of the Universe was less than a Gyr, that the reionization of the hydrogen component of the intergalactic medium (IGM) was substantially completed \citep{Fan06}.  Hundreds of $z\sim6$ galaxies have now been photometrically selected  (e.g. \citealt{Bouwens07, mclure}), and at least tens of Lyman break galaxies (LBGs) and many more \lya\ emitters (LAEs) are now spectroscopically confirmed (\citealt{dow-hygelund, Vanzella, Stanway07, Kashikawa, Shimasaku, Stark10}).   While there are hints of substantial star formation at even higher redshifts (\citealt{Eyles05,Eyles,Yan06, Simcoe, Ryan-Weber, Dunkley, Bouwens08, Bouwens09, bunker09}), $z\sim6$ remains the earliest epoch for which robust galaxy samples are available.  

At $z\sim6$, galaxies are selected via the Lyman break method (\citealt{Meier, Steidel96}), or because of strong \lya\ emission \citep{Hu02, Hu04, Rhoads, Ajiki04,Westra}.    Both methods select star-forming galaxies, but the two populations differ.    Not all LBGs have \lya\ emission \citep{Shapley03, Stanway07, Vanzella} and LAEs can be missed by LBG surveys because they are faint in the continuum or their \lya\ emission contaminates broad-band photometry.    Studies of LAEs are an important complement to LBG surveys, because the faint continuum luminosities and high specific star formation rates of LAEs make 
them likely candidates for galaxy building blocks.

To date, narrow-band imaging surveys have been very successful in finding LAEs at $z\sim3-6$.   Mulitwavelength observations have shown that these LAEs are typically (but not always) younger and less massive, and they have lower dust extinction than LBGs \citep{Finkelstein07, Finkelstein08, Finkelstein09, Gawiser06, Gawiser07, Nilsson07, Lai, Kornei}. 
In addition, \lya\ luminosity functions measured at $z\sim3-6$, show little or no evolution with redshift 
 \citep{Shimasaku, Gronwall, Dawson, Murayama, Ouchi08}.   At higher redshifts measurements of the \lya\ luminosity function may allow a determination of the neutral 
 hydrogen fraction in the intergalactic medium and LAE clustering could help constrain models of the patchiness of of the reionization process   \citep{Santos04, Santos+04, HC05, Furlanetto, Dijkstra07, mcquinn}.  In fact, \cite{Kashikawa} now report a measured decline in the \lya\ luminosity function at $z=6.5$, although this is not seen in a compilation of different data sets presented by \cite{MR04}.

Spectroscopic searches for LAEs have the potential to detect  objects with fainter line emission than purely narrow-band imaging,  because they sample with a resolution closer to the intrinsic width of the emission line. This is especially important at $z\ga6$, where observing the faint end of the galaxy luminosity function is challenging. However, the task is challenging and first efforts at blank sky surveys using multi-slit narrowband spectroscopy at Keck Observatory and the Very Large Telescope (VLT) found only low and intermediate redshift interlopers \citep{Tran, MS04}.  More recently, different spectroscopic strategies have met success.    \cite{Sawicki} have found several faint serendipitous LAEs at $z= 4.4-4.9$  in a search of 20\% of the DEEP2 database.   While the area  subtended by DEEP2 slits is only 21.6 arcmin$^2$, a large line-of-sight volume is sampled within in the relatively OH-free portion of the sky spectrum  at $\lambda < 7000$ \AA.  In addition, 
\cite{Rauch08} found 27 $z\sim3$ LAEs in a single longslit spectrum; their $\sim$ 100 hour VLT observation reached a flux  one to two orders of magnitude deeper than most other LAE searches.  
At higher redshifts, slitless spectroscopy with {\it Hubble Space Telescope} (HST) has uncovered a few sources at $z>5$ \citep{Pirzkal}, and longslit spectroscopy of the critical lines of strong gravitational lenses has discovered low-luminosity LAEs at $z\sim6$ \citep{Santos+04}, as well as a few candidates at $z>8$ \citep{Stark07}.

Recently, wider areas have become accessible with multislit narrowband spectroscopy, and LAEs have been found at $z\approx 5.7$.    Using the Inamori Magellan Areal Camera and Spectrograph (IMACS; \citealt{Dressler}), \cite{Martin08} carried out a blind spectroscopic search in 200 arcmin$^2$  in the Cosmic Evolution Survey (COSMOS; \citealt{cosmos}) and the 15 hour field of the Las Campanas Infrared Survey (LCIRS; \citealt{Marzke}).     These observations reach a sensitivity fainter than $L^*$(\lya) at $z\approx5.7$ ($\sim$ several $\times 10^{42}$ erg s$^{-1}$, as measured by \citealt{Shimasaku} and \citealt{Ouchi08}),  and three LAE candidates at  $z\approx5.7$ were confirmed with further spectroscopy.  Following this success, we have carried out a deeper survey in the same fields (Dressler et al.,  in prep.), uncovering at least several sources at luminosities that were previously only reached with strong lensing searches \citep{Santos+04}.  
    
Of the three sources found by \cite{Martin08}, none are detected in the available continuum imaging, and only one has been previously detected in COSMOS narrow-band imaging  (no narrow-band imaging is available in the 15H field).  In this paper, we present new imaging of these LAEs from the Near Infrared Camera and Multi-Object Spectrometer (NICMOS; \citealt{Thompson_nicmos}). These images, obtained through the \Jnic\ filter with the NIC2 camera, confirm the presence of the two LAEs in the 15H field.  In \S 2 we present our observations and data reduction, as well as the process by which we match our spectroscopically discovered sources with NICMOS detections.  In \S 3 we discuss the properties of these three LAEs, and \S 4 contains a summary of our results.
We use $H_0 = 70 ~{\rm   km s^{-1}~ Mpc^{-1}}$, $\Omega_{\Lambda} = 0.7$, $\Omega_M = 0.3$, and AB magnitudes throughout.

  \begin{figure*}[!ht]
\plotone{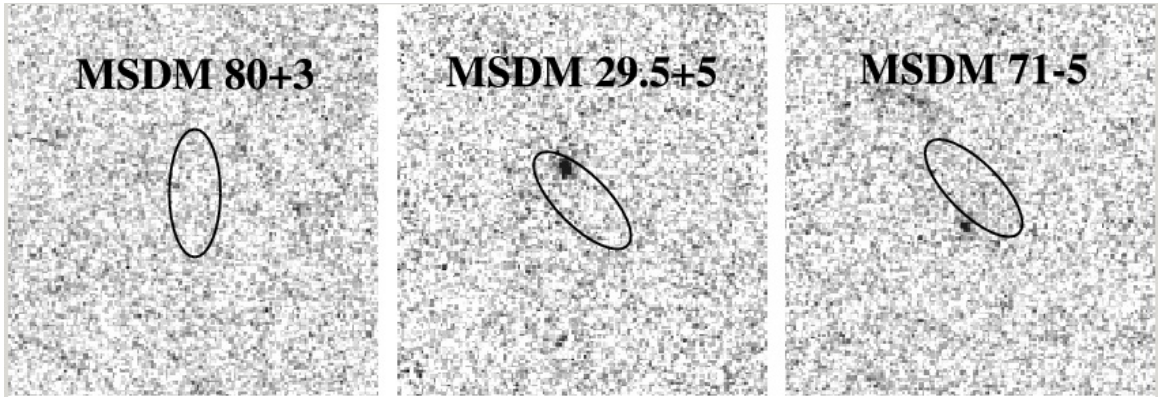}
\caption{Postage stamp images of our NICMOS \Jnic\ observations of the three LAEs.   
 The images are 6\arcsec\ on a side with north up and east to the left. Ellipses mark the predicted positions of the LAEs, assuming that each fell at the center of the Magellan/IMACS spectroscopic slit.  The sizes of the error ellipses are  
 prescribed by the position uncertainties given in \S \ref{offsets}.   
 Although MSDM 80+3 was detected in narrowband observations \citep{Murayama}, the position shown here is-- as with the other two objects-- the one determined from our IMACS spectroscopic observations. 
 Measured positions of the \Jnic\ detections of MSDM 29.5+5 and MSDM 71-5 are given in Table \ref{phot_table}.     }
\label{stampfig}
\end{figure*}

\section{Observations and Data Reduction}

\subsection{NICMOS Imaging}
\label{obs}
Each of the three LAEs were observed for 5 orbits (13.4 ks)  with the NICMOS 2 camera.  The images were then processed in several steps to make final mosaics.  First,  images were corrected for the quadrant-dependent variable bias (the ``pedestal effect''), and South Atlantic Anomaly (SAA) darks were subtracted from impacted orbits.  Next, we used the IRAF task {\it rnlincor} to correct for the count-rate dependent non-linearity documented by \cite{deJong}.  With this approach, the zero point remained unchanged at \Jnic\ = 23.69.    For our configuration of \Jnic\ with NICMOS 2, the non-linearity correction was substantial, and amounted to $-0.25$ mag for the \Jnic\ = 23 - 26 magnitude sources in our images.   Following this correction, the images were sky subtracted, using the NICRED package \citep{nicred}. Then, the remaining vertical and horizontal bands in the images were removed by subtracting a model sky frame constructed by taking the median of each column and smoothing it by a three pixel wide boxcar.  (This step was repeated for rows to remove any top-to-bottom banding or gradients.)  After this, we identified bad pixels in the images following the procedures used by the  Multidrizzle software \citep{multidrizzle}:  A truth image was made by aligning the images and creating a median stack.  This image was shifted back to the frame of the original input image, and pixels deviant by more than 3 $\sigma$ were flagged as bad pixels.  Finally, images were combined with drizzle with bad pixels masked using parameters recommended in the Dither Handbook: pixfrac = 0.6, and scale = 0.5.  The final output pixels are 0.038\arcsec. 

The resulting PSF, measured from the two point sources in these images has a FWHM of 0.1\arcsec.  Sensitivity was measured by randomly placing 0.6\arcsec\ diameter apertures in the image, rejecting those that fell on sources, and fitting a Gaussian to the "counts-per-aperture"  distribution, as described in \cite{thesis}.   We found a 5$\sigma$ sensitivity of \Jnic\ = 26.5 in this aperture.   Using the point sources in our images, we measured the aperture correction to be 0.08 magnitudes, so the 5$\sigma$ total sensitivity is \Jnic\ =  26.4 (for a point source).  Finally, using a few common sources per image, the astrometric solutions to the NICMOS images were aligned with  the COSMOS survey for the 10h field, and the SDSS for the 15h field.     

\subsection{Identifying the LAEs in NICMOS images}
\label{offsets}
The spectroscopic search  and confirmation data are presented in \cite{Martin08}, where a difference was discovered between the World Coordinate System (WCS) zero-points of COSMOS and that of our data.   Namely, the position of MSDM 80+3 in the COSMOS/Subaru narrow-band imaging is nearly 1\arcsec\ west 
of the position found by Martin et al.   In light of this offset, we improved our technique for deriving 
coordinates, by mapping a WCS solution to images taken through our slitmask.   We 
tested these solutions for both the COSMOS and 15H field, using dozens of objects (per field) that have continuum detections in our spectra.  We  found  (1$\sigma$) position uncertainties of 0\farcs3-0\farcs5 along the slits, and 1\farcs0 perpendicular to the slits.  The latter uncertainty naturally arises in all 
blind spectroscopic searches, because sources are not necessarily centered in the slits.    For the COSMOS field, the slits are oriented E-W, so these
uncertainties correspond to RA and Dec, respectively. For the 15H field the slit PA was -45\degr.

In Figure \ref{stampfig}, we show the NICMOS \Jnic\ images at the positions of the three LAEs.  No obvious source is detected at  the position of MSDM 80+3, but sources are detected  within the 1$\sigma$ position error ellipse for MSDM 29.5+5 and just outside it for MSDM 71-5.    Photometry is presented in Table \ref{phot_table}.    We used 
0.6\arcsec\ diameter apertures, centered on the positions of the sources shown in Figure \ref{stampfig} for MSDM 29.5+5 and MSDM 71-5, and at the position of the narrowband detection for MSDM 80+3 (see Figure \ref{stamp80+3}).  A point source model was adopted for the aperture correction, as explained in \S \ref{obs}.   While more light will be missed from extended sources, this loss amounts to at most 0.1 to 0.3 magnitudes (in addition to the point source aperture correction), as nearly all $z\sim6$ galaxies have half-light radii, $r_{hl} < 0\farcs2$ \citep{Bouwens04, Ferguson}.  
As we cannot accurately determine the amount of missed flux, we adopt the point source aperture correction of 0.1 magnitudes, and note this systematic uncertainty.

In order to determine whether the sources are likely to be associated with the LAEs, we 
calculated the probability that the NICMOS detections in the 15H field are from foreground galaxies.  To do this we obtained the publicly available catalogs of the NICMOS Ultra-Deep Field\footnotemark[5] \citep{Thompson_udf}\footnotetext[5]{http://archive.stsci.edu/prepds/udf/udf\_hlsp.html}.
%Using photometry in 0.6\arcsec\ diameter apertures for consistency with our aperture photometry of the LAEs, we find cumulative number counts of $\sim$60 arcmin$^{-2}$ down to \Jnic\ = 26.1 (the aperture magnitude of the fainter of the detected sources). 
For consistency with our aperture photometry of the LAEs, we used 0\farcs6 diameter apertures.  To \Jnic\ = 26.1, which is the aperture magnitude of the fainter 
of the two detected sources, we found cumulative number counts of $\sim$60 arcmin$^{-2}$.  
Therefore, there is less than a one per-cent chance of a foreground interloper falling within our positional error ellipse.  For the brighter source, this probability is about 50\% lower.  We conclude with greater than $99$\% confidence that  each of the NICMOS detected sources is indeed an LAE discovered by \cite{Martin08}.

\subsection{\lya\ contribution to broadband photometry?} 
%Contribution to broadband photometry from strong \lya\ emission is possible in some cases \citep{SdB10}, and 
%in our case \lya\ is included in the 6000\AA\ wide \Jnic\ bandpass. 
Strong \lya\ emission can in some cases contribute appreciably to broadband flux density \citep{SdB10}, and for our objects this could happen.  At $z=5.7$, \lya\ is included in the wide \Jnic\ bandpass, which covers approximately 0.8 to 1.4 \micron.  
However, the contribution is negligible for our objects-- our measured \lya\ fluxes account for  only 5-10\% of the \Jnic\ flux densities  for the two detected sources.  This enhancement lies within our 
uncertainties, so no corrections are made to our photometry and equivalent width measurements.

\section{Discussion}  
\subsection{MSDM 80+3: Comparison to Narrowband Observations}
\begin{figure}[!t]
\plotone{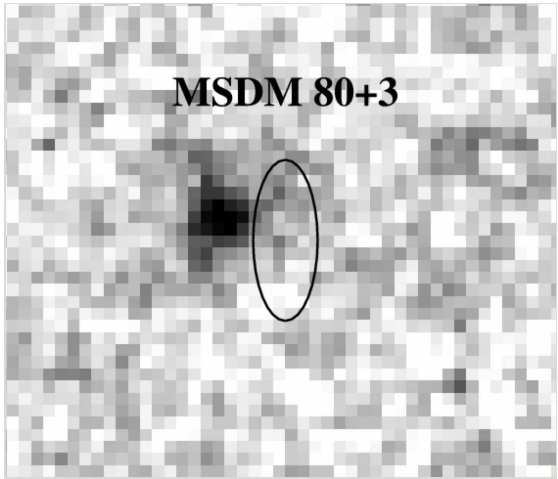}
\caption{NB816 image of MSDM 80+3, 6\arcsec\ on a side with north up and east to the left.  The ellipse marks the predicted position, at the center of the slit, with its size prescribed by the position uncertainties described in \S \ref{offsets}.  In this image, our IMACS slit is located left to right.  The image shown here is the original resolution (0.7\arcsec\ seeing), rather than the PSF matched image used for photometry.} 
\label{stamp80+3}
\end{figure}

These NICMOS observations provide the first direct images of two sources (MSDM 29.5+5 and MSDM 71-5) which had previously only been detected in our spectroscopic data.     On the other hand, the source which is undetected in \Jnic, MSDM 80+3,  shows a strong detection in Subaru narrowband imaging with the NB816 filter (object \#55 in \citealt{Murayama}). In Figure \ref{stamp80+3}  we show the narrowband image of MSDM 80+3, with our position error ellipse overlaid.   Our improved coordinates for MSDM 80+3 are roughly consistent with the NB816 source in Murayama et al.  The newly derived position is 0\farcs7 {\it west} of the NB816 detection, while the previous position given in Martin et al. was 1\arcsec\ {\it east}.   This position offset along the slit is still large compared to other sources, including those discussed in \S 2.2 and for 52 lower redshift line emitters that are in common between Martin et al. and the COSMOS NB816 sample.  
Given the slight offset perpendicular to the slit (0\farcs4) shown in Figure \ref{stamp80+3}, one possibility 
is that the \lya\ emission from this object is extended, and our slit subtended an outlying region of the galaxy.   Some evidence for this hypothesis is present in the Subaru narrowband image (Figure \ref{stamp80+3}), where MSDM 80+3 is marginally resolved.     

The absence of continuum emission could indicate that the source is extended in the continuum, although extended \lya\ emission need not imply a similarly extended stellar continuum.   The upper bound of $i-$ dropout LBG sizes is $r_{hl}  < $ 0\farcs2 (in the rest-frame UV; \citealt{Bouwens04}), which corresponds to 5.3 pixels in our drizzled NIC2 images.   Using the IRAF artdata package, we verified that a galaxy of this size would be undetected in our NICMOS data at  \Jnic\  $> 26.5$. %, in other words, fainter than $L^{*}_{UV}$.   
It is tempting to conclude from the low-surface brightness signal at the position
of MSDM 80+3 in our NICMOS image that it may be more extended in the rest-frame UV.    However, this LAE fell on the 
bad  central column\footnotemark[6] \footnotetext[6]{http://www.stsci.edu/hst/nicmos/performance/anomalies} in these observations, and similar artifacts appear at other locations along the column. The weight maps produced by drizzle indicate that noise is $\sim 30$\% higher at these locations relative to the rest of the image, so the 
 extended feature is not significant.  Nevertheless, the \Jnic\ non-detection could imply that MSDM 80+3 is more extended in the continuum than the other two LAEs 
 (MSDM 29.5+5 and 71-5), which have half-light-radii (measured with SExtractor) of 0\farcs15 and 0\farcs14.

The hypothesis that the \lya\ emission may be extended is also supported by our line flux measurement, 
which is a factor of two lower than the Subaru NB816 imaged line flux.  This difference is larger than is expected for a point source, unless 
the object fell on the edge of our 1\farcs5-wide slit.   As noted above, in the narrow-band image coordinates, MSDM 80+3 is 0\farcs4 north of the slit-center.   Using the point source
slit loss models from Martin et al. (2008), at this position we expect only about a 20\% slit loss.  
The effect of the non-uniform response of the NB816 filter on SuprimeCam is negligible for MSDM 80+3, because our measured wavelength is near  the central wavelength 
of the NB816 transmission. Furthermore, lines at wavelengths away from the 
center of the NB816 bandpass would have decreased observed \lya\ flux relative to our spectroscopic measurement-- the opposite of what we observe.
Generally, sources found through blind spectroscopy do not fall at the center of the slit; for larger samples the impact of slit-losses on the luminosity function
is modeled statistically.

\subsection{Properties}
Equivalent widths, UV luminosities, and UV continuum derived star formation rates  are given in Table \ref{phot_table}. 
We assume a flat UV spectrum ($f_{\lambda} \propto \lambda^{-2}$), consistent with observations of LAEs with \lya\ luminosities and redshifts similar to those of our objects \citep{Pirzkal}, and with $z\sim6$ LBGs in the NICMOS Ultra Deep Field \citep{Stanway05, Bouwens_uvslope}.   Therefore, although the center of the \Jnic\ filter corresponds to $\sim 1700$  \AA\ in the rest frame, our best estimate is that k-corrections are negligible.  We note that $M^*_{UV} = -20.2$ at $z\sim6$ \citep{Bouwens07}, so the UV-luminosities that we measure are consistent with typical sources observed at this redshift-- within a factor of two of $L^*$ for the two \Jnic--detected sources.

 Star formation rates (SFRs)  are derived using the 1500 \AA\ conversion from \cite{MPD98}.  UV-continuum based estimates of
  star-formation rates are thought to be more reliable than those from \lya; nevertheless, these continuum-based estimates come with several caveats.  
   First, the conversion is valid only when the characteristic stellar age is older than 
 the main sequence lifetime of O- and B-stars.  At young ages ($\la 20$ Myrs), the true SFR may be a few times larger.  
 Second, we have not corrected the UV-luminosities for dust extinction, so  again, the SFRs given in Table \ref{phot_table} may be underestimated.   
   However, in general, $z\sim6$ galaxies are 
 not thought to be very dusty \citep{Bouwens_uvslope}, so dust corrections to our SFRs need not be very large.    Third, uncertainties in the IMF slope and mass cutoffs may be just as important, as they can
 change the SFR by a factor of a few \citep{thesis}.   On the other hand, the two-photon nebular continuum is not expected to make a significant contribution to the 
 UV-luminosity, since it is generally dwarfed by the stellar contribution.   While some sources are purported to have spectra dominated by the two-photon continuum,
 such objects appear to be rare \citep{Fosbury03, Raiter}, and are predicted to have rest-frame \lya\ equivalent widths $\ga 1000$ \AA\ (\citealt{Schaerer02}, with coefficients from \citealt{Aller}).

\begin{figure}
\plotone{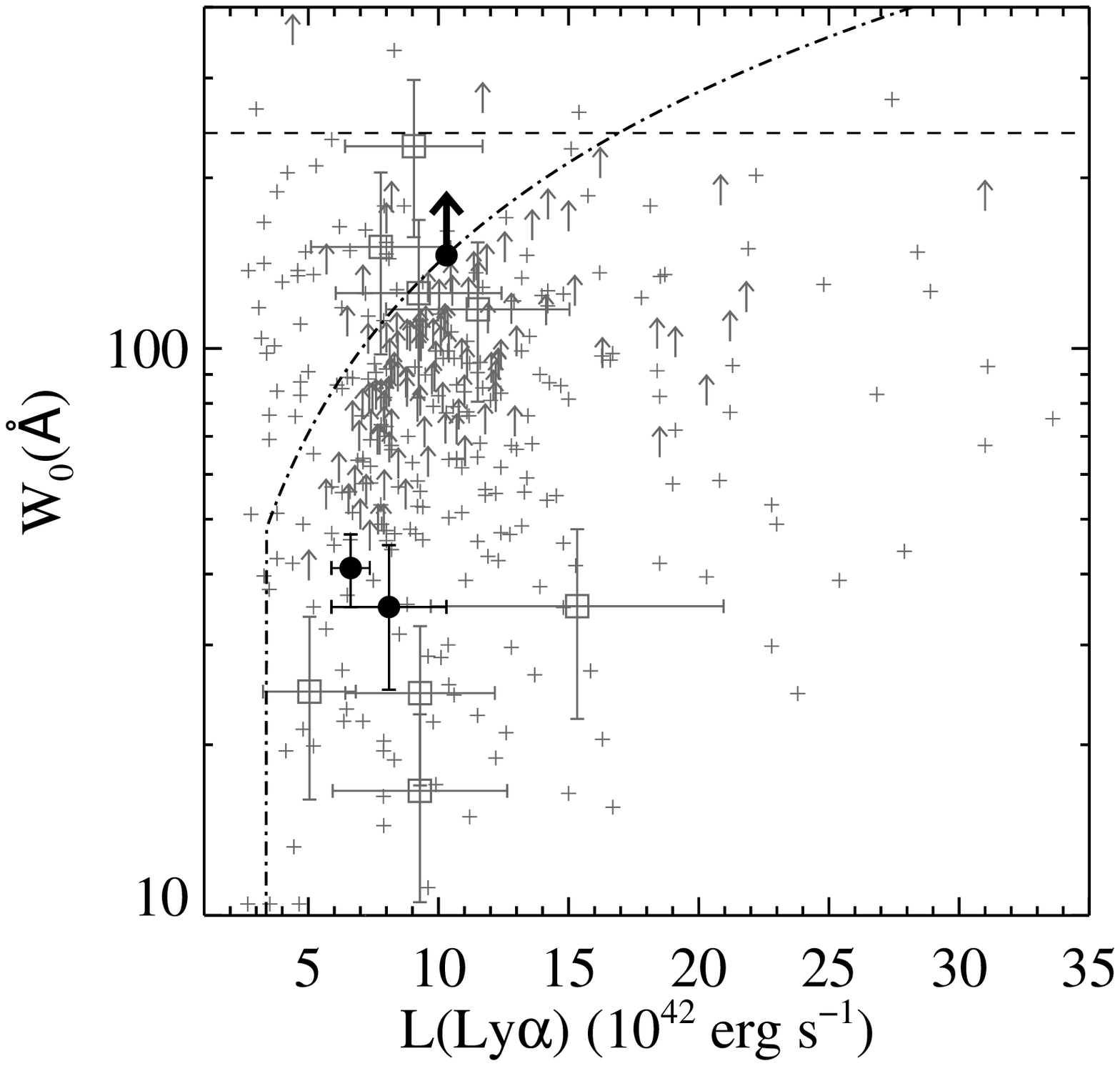}
\caption{Rest frame equivalent widths of the sources in our survey (black) fall in the normal range 
established by various other \lya\ searches at $z=4-6$ (grey).  These include sources discovered with
the ACS grism (squares; \citealt{Pirzkal}) and narrowband imaging surveys  \citep{Hu04,Shimasaku, Murayama, Ouchi09, Wang}.  The dot-dashed line shows the maximum detectable equivalent width when  $M_{UV} = -19.8$ is reached, as is the case with our NICMOS observations.  The dashed horizontal line marks $W_{0} = 240$ \AA, which is often quoted as the maximum equivalent width seen in ``normal'' stellar populations \citep{CF93, MR04}.   %While many equivalent widths are lower limits,  and may in fact exceed this limit, only a few sourc  
It is important to note that the equivalent width uncertainties typically  exceed 100\AA\ when $W_{0}$ is large, so the sources above this line should be viewed with caution.  }
\label{ewplot}
\end{figure}

 The equivalent widths that we measure are consistent with those of LAEs at high redshift, as shown in Figure \ref{ewplot}.     We do not make any correction for IGM absorption on the \lya\ emission, as has been done  for some objects at $z\sim6$ (e.g.\ \citealt{Shimasaku}).  While it is possible that the \lya\ emission emergent from the galaxy is larger than observed, the amount of attenuation is uncertain because the resonant scattering of \lya\ photons shifts the emission towards redder wavelengths \citep{Verhamme}.    It is not surprising that we do not observe any sources that are both UV luminous  and have high-equivalent widths.  As \cite{Nilsson} point out, this may be a consequence of both classes of objects being rare, rather than a correlation of equivalent width with UV luminosity as has been suggested by \cite{Ando}.

 The three LAEs that we present are consistent with the range of equivalent widths that are easily explained by ``normal''  stellar populations, with ages older than 10 Myrs and a Salpeter IMF.    However, deeper continuum observations of MSDM 80+3 could prove that its equivalent width is much larger.    There are several different models that can explain very large equivalent widths.  First,  extremely young ages and/or top-heavy initial mass functions (IMFs)  can result in rest-frame equivalent width $W_{0} > 240$ \AA\ (e.g.\ \citealt{MR04}).   Second, a multi-phase  interstellar medium has sometimes been invoked to
 preferentially absorb UV continuum photons (\citealt{Neufeld, HO06, Scarlata, Finkelstein09}); and third, gravitational cooling radiation in the absence of star formation is predicted to result in $W_{0} > 1000$ \AA\  (\citealt{Dijkstra09}).   However, such high equivalent width objects are probably rare, as only  a few sources in Figure \ref{ewplot} lie at $W_{0} > 240$ \AA.  It is important to note that
 the uncertainty on such high equivalent width sources typically exceeds 100\AA\ because of weak continuum detections, so these measurements should be viewed with caution.

It is interesting to consider whether the LAEs in our survey would be selected as $i-$ dropout LBGs at $z\sim6$, because the connection between the two populations is currently unclear.   While it is generally understood that LAEs can be missed by LBG searches when  their continuum is too faint, the selection of brighter sources can also be influenced by strong line emission (e.g.\ \citealt{Stanway07, Stanway08_contam}).   At $z\le 5.7$, \lya\ falls in the $i$ -band, and galaxies can sometimes be  too blue in $i-z$ to be included in $i-$dropout samples, even when an identical galaxy without \lya\ emission would be included in such a sample.   (It is worth noting that such objects could be included in $z\sim5$ R-dropout samples, provided that the $i-z$ color is not too red.)
Strong  \lya\ emission may influence the color selection of all three galaxies presented here.  Assuming a flat UV slope ($f_{\lambda} \propto \lambda^{-2}$), we estimate $i-z = 1.2$ and 1.3 for MSDM 29.5 and MSDM 71-5, near the $i-z = 1.3$ cut used by most LBG studies (e.g. \citealt{Bouwens07}).  The inclusion of sources such as these would depend on photometric scatter and the true value of the UV slope.  On the other hand, MSDM 80+3, which is undetected in the \Jnic\ images, would be unlikely to be included in $i-$dropout samples.  The \lya\ flux contributes substantially to the $i-$band, so 
even in the absence of any continuum we would expect $i = 27.6$.  Assuming \zacs $> 26.9$-- as inferred from our \Jnic\ limit-- implies that we would expect $i-z \le 0.7$.  Therefore, even if MSDM  80+3 could be detected in very deep imaging such as the UDF, it would be much too blue to be included in the $i-$ dropout samples.   This exercise emphasizes the importance of including \lya\ emission in the modeling of LBG selection functions, as is done by \cite{Bouwens07}. 

\section{Conclusions} 
Spectroscopic searches for high-redshift emission-line galaxies have now been shown as a viable technique for improving
sensitivity to faint emission line galaxies \citep{Santos+04, Sawicki, Lemaux}.  In particular, \cite{Martin08} have successfully identified three $z=5.7$ LAEs with multislit narrowband spectroscopy.   We have presented NICMOS observations of these three LAEs, obtained with the high spatial resolution NIC2 camera.  These images confirm the two sources which had only previously been seen in our IMACS spectroscopic observations.   Furthermore, the \Jnic\ data provide some constraints on the properties of these galaxies.  We find that the UV luminosities, star formation rates, equivalent widths and sizes are all consistent with a redshift of $z=5.7$.   The two detected sources have $L_{UV}$ within a factor of two of $L^*_{UV}$ (as measured by \citealt{Bouwens07}), and equivalent widths are well within the range of values seen at all redshifts.   In addition,   the non-detection of MSDM 80+3, in combination with a substantial slit-loss of \lya\ flux suggests that
the source may be somewhat extended.  

These new observations further prove the success of multislit narrowband spectroscopy as a means to uncover faint, 
high-redshift emission line galaxies.  Further efforts for extending this technique to much fainter LAEs are already underway (Dressler et al., in prep).     Longer exposures with the recently upgraded IMACS detectors can reach \lya\ luminosities at $z\sim6$ that have previously only been realized in gravitational lensing surveys along the critical lines of galaxy clusters (e.g.\ \citealt{Santos+04}).   At the limit of our new survey ( at least $4\times 10^{-18} ~{\rm  erg ~s^{-1} ~cm^{-2}} $),   rest-frame equivalent widths above 150 \AA\   would imply a $z-$band magnitude fainter than 29- a regime that is currently only accessible from the small area covered by the UDF and UDF Parallels.    These  new blind spectroscopic observations will provide an ideal sample of low-luminosity $z=5.7$ LAEs 
that will be detectable with NIRCAM, NIRSPEC,  and the Tunable Filter Imager on the  {\it James Webb Space Telescope}   in modest  exposure times  ($\la$ 10,000 seconds).   
Such future observations will provide critical insight into the building blocks of galaxies by measuring their stellar and nebular (line+continuum) emission.

\begin{deluxetable*}{lcrrccccc}
\tabletypesize{\scriptsize}
\setlength{\tabcolsep}{0.05in} 
\tablecolumns{9}
\tablehead{
\colhead{Name} & \colhead{RA} & \colhead{Dec} & \Jnic\ & F(\lya) & $W_0$ & \colhead{$M_{UV}$} & \colhead{SFR} & \colhead{ $r_{hl}$  } \\
\colhead{}  & \colhead{(J2000) } & \colhead{(J2000)}  & \colhead{(AB)}  &  \colhead{($ 10^{-18}$ erg s$^{-1}$ cm$^{-2}$)} & \colhead{(\AA)} & \colhead{(AB)} & \colhead{ ( $M_{\sun}$ yr$^{-1}$ )} & \colhead{ (\arcsec)}  } 
\startdata 
MSDM 80+3    & 10 00 30.413	  & 02 17 14.81  & $>26.9$ & 28 $\pm4$\tablenotemark{a}& $>146$ & $> -19.8 $ & $< 4.0$ & \nodata \\ 
MSDM 29.5+5 & 15 22 57.900    & -00 07 36.80 & 26.0 $\pm$ 0.1 & 18 $\pm$ 2  & 41 $\pm6$ & -20.7& 10   & 0.15 \\  
MSDM 71-5     &    15 24 08.920  & -00 10 43.31  & 25.6 $\pm$ 0.1  & 22 $\pm$ 6 & 35 $\pm 10$ & -21.1 & 14  & 0.14 \\
\enddata
\tablecomments{\Jnic\ magnitudes are total, and equivalent width is given in the rest-frame.   Limits are 3$\sigma$, total, for a point source.  As discussed in \S  3.1 the limit on MSDM 80+3 will be weaker if it is significantly extended.  Coordinates given are measured from direct images of the sources:  NICMOS for MSDM 29.5+5 and MSDM 71-5, and Subaru NB816 for MSDM 80+3.  Star formation rates are estimated from the \cite{MPD98} conversion, but will be higher if extinction is large or the galaxies are extremely young.   }
\tablenotetext{a}{Taken from the narrow-band imaging measurement \citep{Murayama}.  }
\label{phot_table}
\end{deluxetable*}

\acknowledgements
This work is supported by HST GO-11183.  We would like to thank Matt Auger, Peter Capak, and Kristian Finlator for helpful suggestions and discussions. 
We are grateful to the anonymous referee for comments that improved this manuscript.

\end{document}